\documentclass[amsmath,amssymb,reprint,onecolumn]{revtex4-1}
\usepackage{graphicx,natbib,overpic,xcolor,bm}
\begin{document}

%\title{The initial symmetry breaking in convective self-aggregation}
\title{How weakened cold pools open for convective self-aggregation}
%\title{Cold pool interaction hampering the initiation of convective self-aggregation}

\author{Silas Boye Nissen}
\email{silas@nbi.ku.dk}
\affiliation{Niels Bohr Institute, University of Copenhagen, Blegdamsvej 17, 2100
Copenhagen, Denmark.}

\author{Jan O.~Haerter}
\email{haerter@nbi.ku.dk}
\affiliation{Niels Bohr Institute, University of Copenhagen, Blegdamsvej 17, 2100 Copenhagen, Denmark. \\ 
Physics and Earth Sciences, Jacobs University Bremen, Campus Ring 1, 28759 Bremen, Germany. \\ 
Complexity and Climate, Leibniz Center for Tropical Marine Research, Fahrenheitstrasse 6, 28359 Bremen, Germany.}

\date{\today}

\begin{abstract} %%% MAX 150 WORDS
\noindent
In radiative-convective equilibrium (RCE) simulations, convective self-aggregation (CSA) is the spontaneous organization into segregated cloudy and cloud-free regions. 
Evidence exists for how CSA is stabilized, but how it arises favorably on large domains is not settled. 
Using large-eddy simulations (LES), we link the spatial organization emerging from the interaction of cold pools (CPs) to CSA. 
We systematically weaken simulated rain evaporation to reduce maximal CP radii, $R_{\text{max}}$, and find reducing $R_{\text{max}}$ causes CSA to occur earlier. 
We further identify a typical rain cell generation time and a minimum radius, $R_{\text{min}}$, around a given rain cell, within which the formation of subsequent rain cells is suppressed. 
Incorporating $R_{\text{min}}$ and $R_{\text{max}}$, we propose a toy model that captures how CSA arises earlier on large domains:
when two CPs of radii $r_{i,j}\in[R_{\text{min}},R_{\text{max}}]$ collide, they form a new convective event. 
These findings imply that interactions between CPs may explain the initial stages of CSA.% symmetry-breaking and that larger domain size facilitates self-aggregation.
\end{abstract}

\maketitle

\section{Key Points} %%% MAX 140 CHARACTERS INCLUDING SPACES EACH
1. Smaller cold pool radii in large-eddy simulations (LES) diminishes the time to reach convective self-aggregation.

2. Incorporating the maximal cold pool radius and the radius of suppressed regions into a simple model gives realistic self-aggregation.

3. The model supports the role of cold pools and captures the effect of domain size, where larger domain size facilitates self-aggregation.

\section{Plain Language Summary} %%% MAX 200 WORDS
Convective self-aggregation (CSA) describes the emergence of persistently dry, cloud-free areas in numerical simulations. It has been suggested as a possible mechanism for tropical cyclone formation and large-scale events such as the Madden-Julian Oscillation. Some understanding of the persistence of CSA exists. However, how CSA initially emerges remains poorly understood. Recently, the dynamics of cold pools (CPs) have been associated with the organization of convective events. CPs are radially expanding pockets of dense air that form under precipitating thunderstorms. In this work, we ask how weakening CPs could facilitate the emergence of CSA. By analyzing high-resolution numerical simulations, we show that reducing rain evaporation shortens the time before CSA sets in. 
These simulations demonstrate that CPs reach greater radii when rain evaporation is large. In addition, we find that new convective events occur near the point where two CPs collide. Finally, we find a minimum CP radius within which CPs are too negatively buoyant to initialize new convective events. 
Building on these numerical findings, we propose a simple idealized mathematical model that approximates CPs as expanding and colliding circles. 
We show that this model can capture the emergence of CSA. We conclude that the lack of CPs facilitates CSA.

\section{Introduction}
When evaporation of rain from convective clouds is strong, so is the associated sub-cloud cooling and density increase \cite{simpson1980downdrafts,engerer2008surface}, forcing the resulting cold pools (CPs) to spread more quickly and cover larger areas \cite{romps2016sizes,torri2015mechanisms,zuidema2017survey}. 
Such pronounced CP activity has repeatedly been suggested to hamper convective self-aggregation (CSA) in radiative-convective equilibrium (RCE) numerical experiments \cite{jeevanjee2013convective,muller2015favors,holloway2016sensitivity,hohenegger2016coupled,yanase2020new}.
In these simulations, the atmosphere gradually organizes from an initial homogeneous population of convective updrafts into a segregated pattern with strongly convecting regions and dry, precipitation-free regions \cite{held1993radiative,tompkins1998radiative,bretherton2005energy,hohenegger2016coupled,wing2017convective}. 

Generically, CSA is characterized by the appearance of long-lived dry and warm patches, within which rain is suppressed \cite{holloway2017observing}. 
%Once these patches significantly exceed typical spatial and temporal scales of individual rain cells (a few kilometers and hours), the formation of patterns can be referred to as {\it symmetry breaking}.
Further drying increasingly occurs through enhanced radiative cooling in already dry regions and the resulting subsidence. 
Later, the dry regions expand and merge, eventually leaving only one contiguous moist area with intense low-level convergence feeding convection. 
Surface latent and sensible heat fluxes --- which increase under stronger surface wind speed --- may further increase such low-level moisture convergence.

CPs are capable of mediating organization, as they effectively relay "information" between one precipitating cloud and its surroundings. 
Physically, CPs spread as density currents along the surface, carry kinetic energy and buoyancy, and modify the thermodynamic structure near the CP edges \cite{tompkins2001organizationCold,langhans2015origin,de2017cold}. 
CPs thereby act to pattern the convectively unstable atmosphere, establishing connections between the loci where new convective cells emerge and loci at which the previous cells dissipated. 
In particular, new cells were suggested to be spawned by the CP gust front alone or by collisions between mobile gust fronts \cite{de2017cold,glassmeier2017network,cafaro2018characteristics,fuglestvedt2020cold}. 
Inspired by the notion of CP interactions, conceptual work has formulated CPs as cellular automata \cite{boing2016object,windmiller2017organization,haerter2019circling,haerter2020diurnal}, and CP representations have been incorporated into large-scale models \cite{grandpeix2010density}.
%Recent work suggests that in RCE simulations, the interaction between CPs might be more subtle, where CP gust fronts, when colliding, build up a circulation persisting over several hours. 
%The resulting buildup of moisture and temperature at the locus of collision eventually gives rise to a new convective event \cite{fuglestvedt2020cold}.

Studies on convective self-aggregation often argue that sufficiently large domain size is required for the phenomenon to emerge \cite{bretherton2005energy,muller2015favors}, especially when the horizontal resolution is coarse, e.g., at least 500 km for resolutions finer than 2 km \cite{yanase2020new}.
To examine this claim more closely, for deliberately small domain sizes and fine horizontal resolution, we here show that CSA sets in earlier when CPs are weakened through reductions in rain evaporation --- that is, when the CP maximal radius, termed $R_{\text{max}}$, is reduced. 
We track the CP gust fronts to motivate that loci of gust front collisions are preferable for subsequent convective rain cells --- hence, that CP interactions are essential in organizing the precipitation field.
Dependent on rain evaporation, we further detect a minimal distance $R_{\text{min}}$, within which subsequent rain cells are unlikely to form, as well as a typical generation time.
Using these findings, we build, simulate, and analyze a simple mathematical model, which helps understand cloud-free regions' formation.
We explore the phase diagram of this model and find that, when initialized at a high density of rain cells, the population of subsequent rain cells will either remain high or transition to a low state with only a subregion covered by rain cells. The change from high to low density occurs later for large $R_{\text{max}}$ or small domain sizes $L$.
%We explore the phase diagram of this model and find that two phases exist: a disaggregated state and an aggregated state with $R_{\text{max}}$ and $R_{\text{min}}$ as control parameters. 

\begin{figure*}[htb]
\centering
\includegraphics[width=0.7\linewidth]{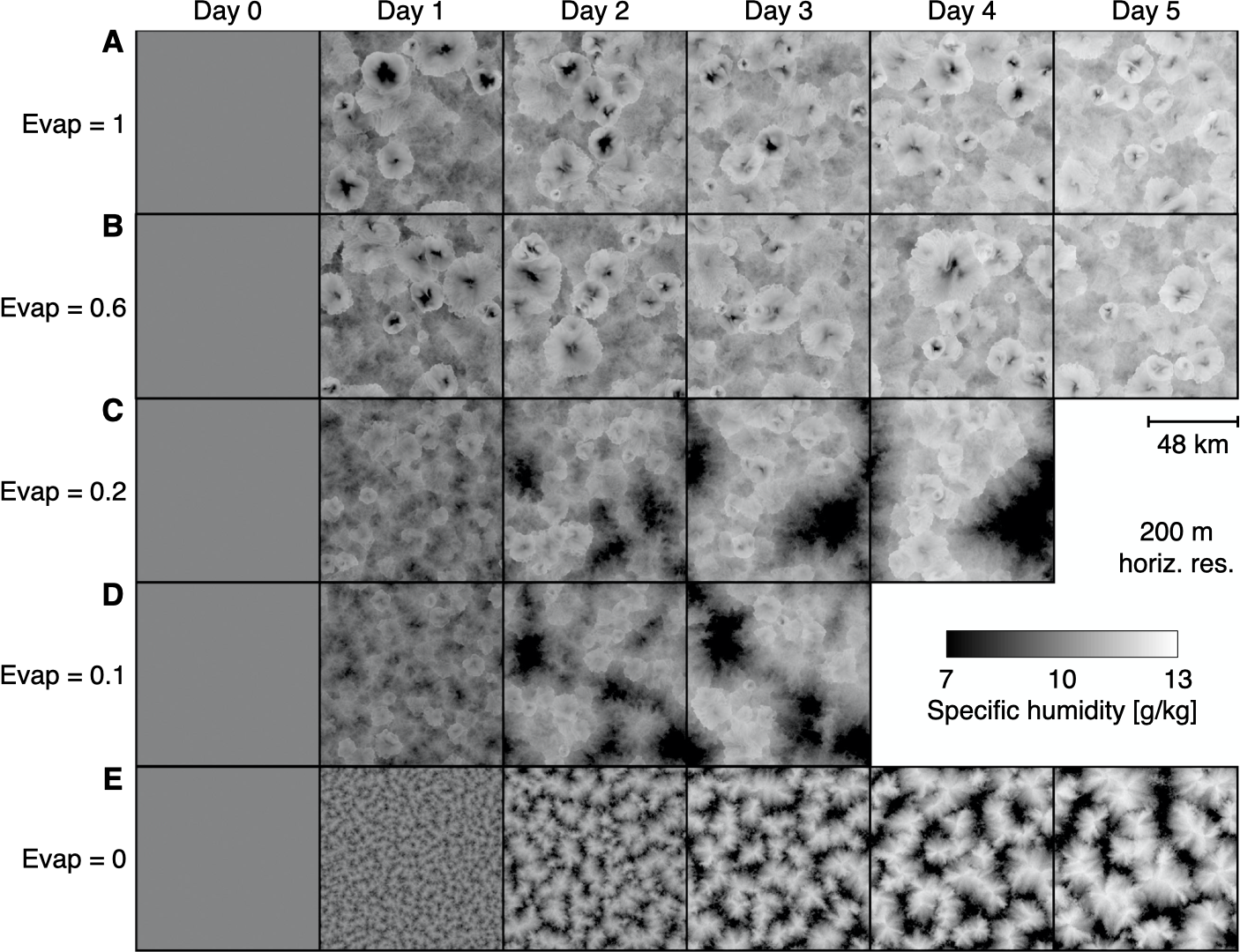}
\caption{{\bf The onset of convective self-aggregation}.
Near-surface specific humidity $q_v$(50 m) the first 10 min of each simulation day in RCE simulations with various degrees of rain evaporation. 
{\bf (A)} Realistic rain evaporation (control simulation). 
{\bf (B)} 60\%, {\bf (C)} 20\%, {\bf (D)} 10\%, and {\bf (E)} 0\% rain evaporation relative to (A). 
Within cold pool centers, note the pronounced moisture reduction in (A--B) and the weakened moisture reduction in (C--D). 
Besides, note the evolving moisture segregation, which is typical of convective self-aggregation (C--D) and moisture coarsening progression (E).
}
\label{fig:specific_humidity}
\end{figure*}

\section{Materials and Methods}\label{sec:methods}
\noindent
{\bf Large-eddy simulations.}\label{sec:les_data}
%\noindent
We carried out (96 km)$^2$ simulations of the convective atmosphere for up to five simulation days using the University of California, Los Angeles (UCLA) Large Eddy Simulator with sub-grid scale turbulence parametrized after Smagorinsky \cite{smagorinsky1963general}.
We combine with a delta four-stream radiation \cite{pincus2009monte} and a two-moment cloud microphysics scheme \cite{stevens2005evaluation}. 
Rain evaporation is accounted for by \citeauthor{seifert2006two} (\citeyear{seifert2006two}).
All five simulations are identical, except that rain evaporation is varied within fractions $\{0,.1,.2,.6,1.0\}$ of its default.
In the following, the corresponding simulations are labeled "Evap=0", "Evap=0.1", etc.
We reduce computed rain evaporation by the corresponding factor --- hence, less hydrometeor mass is converted to water vapor.
Surface temperatures are set constant to 300 K, and insolation is fixed using a constant equatorial zenith angle of 50$^{\circ}$ (constant 655 W m$^{-2}$) \cite{bretherton2005energy}. 
Surface heat fluxes are computed interactively and depend on the vertical temperature and humidity gradients and horizontal wind speed, which is approximated using the Monin-Obukhov similarity theory.
Temperature and humidity are initialized using a prior approximately 3-day spin-up using 400 m horizontal resolution ({\it see} Fig.~S1). %\ref{fig:InitialConditions}).
The horizontal model grid is regular, and periodic boundary conditions are applied in both lateral dimensions. 
Vertically, the model resolution varies from 100 m below 1 km, stretching to 200 m near 6 km and finally 400 m in the upper layers. There are 75 vertical levels.
The Coriolis force and the mean wind were set to zero with weak random initial perturbations added as noise to break complete spatial symmetry. 
At each output time step of 10 min, instantaneous surface precipitation intensity, specific humidity, temperature, liquid water mixing ratio, outgoing long-wave radiation, and 3D velocities are output at various model levels.
To explore resolution effects, we supplemented the simulations described above using Evap=1 by otherwise similar simulations at 1km, 2km, and 4km horizontal resolution, each using $480\times 480$ horizontal grid boxes.

\noindent
{\bf Tracking of cold pools.}\label{sec:cp_tracking}
Cold pool (CP) gust fronts are tracked using tracer particles \cite{haerter2019circling,henneberg2020particle}. 
In any given time step, a sufficient number of tracers, $n_{tr}$, are placed at the edges (8-neighborhood) of an existing surface precipitation patch, which is a spatially contiguous area of rainy pixels (intensity $I>I_0$, with $I_0\equiv$ .5 mm h$^{-1}$). 
The first tracers are placed when the patch is first detected, and this time defines the time origin for each CP ($t$ = 0 in Fig.~\ref{fig:Rmax}B).
If the number of tracers placed is less than a maximal number $n_{tr,max}\equiv$ 100, i.e., $n_{tr}<n_{tr,max}$, then further tracers are placed at the patch edges during the subsequent timesteps until $n_{tr,max}$ is reached. 
If the rain patch disappears before $n_{tr,max}$ is reached, fewer tracers are used for that CP. 
This procedure was found sufficient in reliably tracking the gust fronts. 
Using a simple Euler method, the tracer particles are advected along with the radial velocity component in the lowest model level ($z$ = 50 m) and reliably settle in the gust fronts surrounding each CP.
As this tracking method is implemented to run "offline," it uses only the recorded discrete 10 min output time steps.
In the initial stages of tracking each CP, this sometimes leads to particles having to "catch up" with the gust front ({\it see} Fig.~\ref{fig:Rmax}A, for examples of such CPs). 
However, it is found that tracers consistently settle in the gust front after few time steps due to the strong radial velocity gradient.
To transparently compare CP radii between the different simulations, which differ in time of rainfall onset (Fig.~\ref{fig:Rmax}B), we first determine the time of onset for each simulation, that is, the time when the first surface rain patch with $I>I_0$ occurred. 
We then track all CP gust fronts during 1100 min. This interval was found sufficient to yield significant statistics on the spreading of each CP, but short enough so that not many CP collisions were encountered.
Conversely, to study collision effects (Fig.~\ref{fig:tracers}), we used a late-stage ($\sim$ 4 days after initialization) of the realistic simulation (Evap=1). 
As CPs are space-filling, any new CP inevitably collides with recent CPs in its surroundings.

\noindent
{\bf Mathematical model.}\label{sec:math_model}
We initialize $N_1= L^2/(10R_{\text{min}}^2)$ points --- the subscript indicates "generation one" --- at random locations selected uniformly on a 2D domain of size $L\times L$ with double-periodic boundary conditions. 
All points grow into circles representing CPs with equal and constant radial speed. 
The expanding circles have centers at $\overline{c}_i = [x_i, y_i]$ and increasing radii $r_i$. 
At the collision point, [$x, y$], of two circles belonging to the same generation $g$ and having radii $r_i$ and $r_j$, a new expanding circle belonging to the subsequent generation $g+1$ emerges only if %given that circles $i$ and $j$ are still expanding and that the radii $r_i$ and $r_j$ of circle $i$ and $j$ fulfill
$R_{\text{min}}<r_i,r_j<R_{\text{max}}$. Hence, for two CPs to % be able to
form a new circle, their centers' separation distance needs to be greater than $2R_{\text{min}}$ and less than $2R_{\text{max}}$. 
%However, due to generally asynchronous initiation times, this range will in fact be even smaller. 
Colliding circles continue expanding, and new circles start growing instantaneously; hence, all circles, at all times, expand with equal and constant speed.

We find the collision point by solving
$(x-x_i)^2 + (y-y_i)^2 = (r_i+dr)^2$
and
$(x-x_j)^2 + (y-y_j)^2 = (r_j+dr)^2$, where $dr$ is the distance from the two circles' rims to the collision point. 
Only collisions that fall onto the straight line between the two circle centers are allowed since that is the collision point with the highest momentum, yielding
$y = \frac{x - x_i}{x_j - x_i} (y_j - y_i) + y_i\,.$
These three quadratic equations have three unknowns ($x, y, dr$), and two solutions. 
One of these two solutions can be ruled out because only the positive real solution is relevant to this model.

Similar to the analytical approach in \cite{haerter2019circling}, we do not run the model strictly chronologically. 
To reduce simulation time, we take advantage of the fact that circles belonging to different generations cannot interact since they grow with equal and constant speed. 
Thereby, we calculate all collision points for each generation before proceeding to the next generation. Generally, the last circle in any generation will initiate later than the next generation's first circle. Therefore, when moving to the next generation, we return to the time when the first circle of that generation was seeded.

The list of potential collision points is calculated for each generation and sorted incrementally by $dr$. We update the system by inserting circles at the collision points if $R_{\text{min}}<r_i,r_j<R_{\text{max}}$, and if the subsequent generation does not occupy the collision point. This process is simulated until circle generation number 500 is reached.%continues until the system has reached a final state.

\section{Results}
\noindent
{\bf Weakening cold pools in RCE simulations speeds up the onset of self-aggregation.}
A control simulation with realistic rain evaporation (Fig.~\ref{fig:specific_humidity}A) shows no indication of CSA. 
We check this by computing the inter-quartile specific humidity difference (Fig.~S2A), %\ref{fig:q_timeseries_max-min}A),
finding a weak initial increase when first CPs set in but no further increase over time.
While leaving the total number of rain cells and domain-average rainfall approximately unchanged (Figs.~S2B and S3), %\ref{fig:NoEvents} and \ref{fig:q_timeseries_max-min}B), 
decreasing the rate of rain evaporation (Fig.~\ref{fig:specific_humidity}B--E) yields a monotonic increase in humidity variation (Fig.~S2A) and overall higher near-surface temperature (Fig.~S2C), along with a systematically earlier onset of persistent dry patches, e.g., near day 2 for Evap=0.2 (Fig.~\ref{fig:specific_humidity}C). 
This comparison underlines findings from \citet{jeevanjee2013convective} and \citet{muller2015favors}, who reported that CPs hamper self-aggregation. 
The five experiments highlight that reducing rain evaporation weakens subsidence drying in the center of CPs ({\it compare} dark spots in Fig.~\ref{fig:specific_humidity}A--B vs.~C--D) and visibly reduces CP radii.
We also note that intermediate values of evaporation appear to allow for a band-like aggregation state, where rain cells form a quasi-one-dimensional chain around one of the horizontal dimensions (Fig.~\ref{fig:specific_humidity}C on day 4).
When rain evaporation is entirely removed (Fig.~\ref{fig:specific_humidity}E), any organizing effect through CPs is absent: one is left with a coarsening process akin to reaction-diffusion dynamics, \cite{windmiller2019convection} small impurities gradually merging into larger structures.
The dominant feedback can be ascribed to the net warming in the troposphere induced by deep convective clouds, thus promoting further convective activity in existing precipitation cells' surroundings.

\begin{figure}[htb]
\centering
\includegraphics[width=0.85\linewidth]{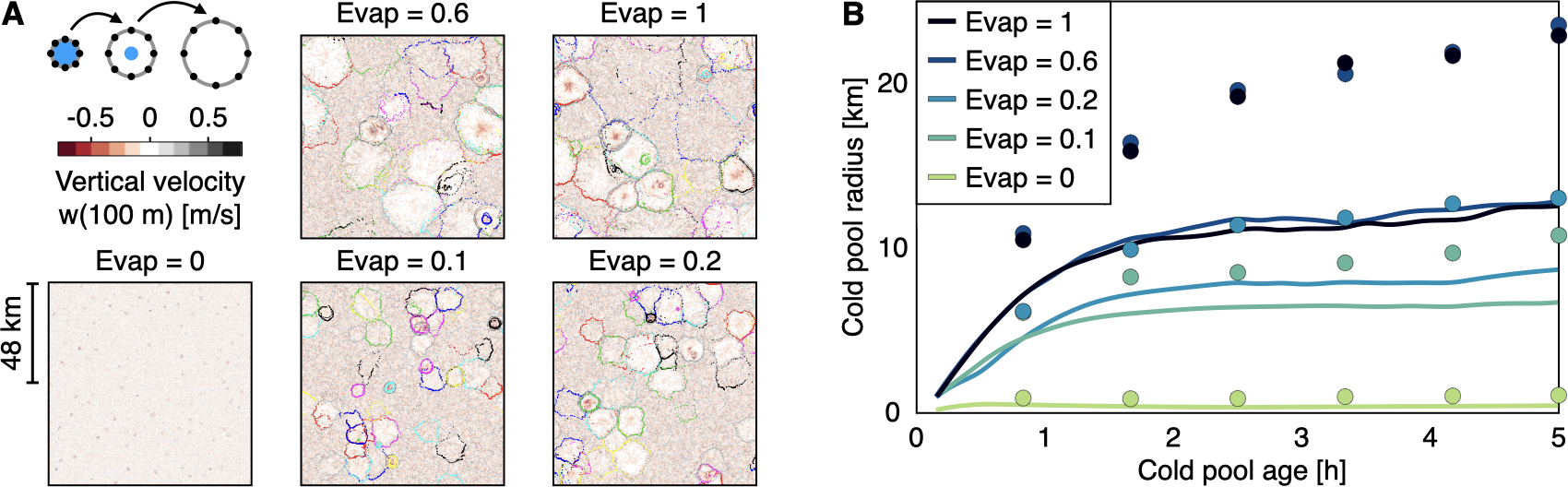}
\caption{{\bf Maximum radius, $\mathbf{R_{max}}$}. 
{\bf (A)} Tracking all cold pool (CP) gust fronts from the onset of precipitation for 1,100 min.
Top-left cartoon: we track a CP gust front (grey rim) by placing tracers (black points) around the rain event (blue spot) and let the tracers move radially away from the rain event with the horizontal wind ({\it Details:} Methods). 
Each panel shows the near-surface vertical velocity field at the end of the time interval and gust front tracers marked by colors indicating different CPs.
{\bf (B)} The average CP radii as the CPs evolve after their emergence (lines) and the 90th radius percentile (dots). 
Note that CPs initially grow quickly but monotonically slow and that maximal CP radii increase with rain evaporation rate.
}
\label{fig:Rmax}
\end{figure}

\noindent
{\bf Measuring cold pool radius.}
Using a rain cell \cite{moseley2019statistical} and CP \cite{haerter2019circling,henneberg2020particle} tracking method, we seed tracer particles at the boundary of surface rain patches ({\it see} cartoon in Fig.~\ref{fig:Rmax}A). We advect these tracers using the radial velocity field, forcing them to gather in pronounced convergence areas caused by the CP gust fronts ({\it Details:} Methods).
Superimposing the resulting pattern of tracers onto the near-surface vertical velocity field (Fig.~\ref{fig:Rmax}A) shows that the tracers indeed gather along the edge of each CP (subsident or featureless vertical wind field).
It is also visually apparent that radii systematically increase with the evaporation rate.
The radius increase is confirmed by plotting the time evolution of the average CP radii in each simulation (Fig.~\ref{fig:Rmax}B).
For Evap=0, radii simply show the radii of the corresponding surface rain cells, as, without CPs, there is no pronounced wind field to advect the tracers. 
For Evap=1, CPs typically expand to $\approx$ 11 km a few hours after initiation, a value comparable to previous simulation results found on various domain sizes \cite{romps2016sizes,tompkins2001organizationCold} and observational findings \cite{black1978mesoscale,zuidema2012trade,feng2015mechanisms}. 

\begin{figure*}[htb]
\centering
\includegraphics[width=0.85\textwidth]{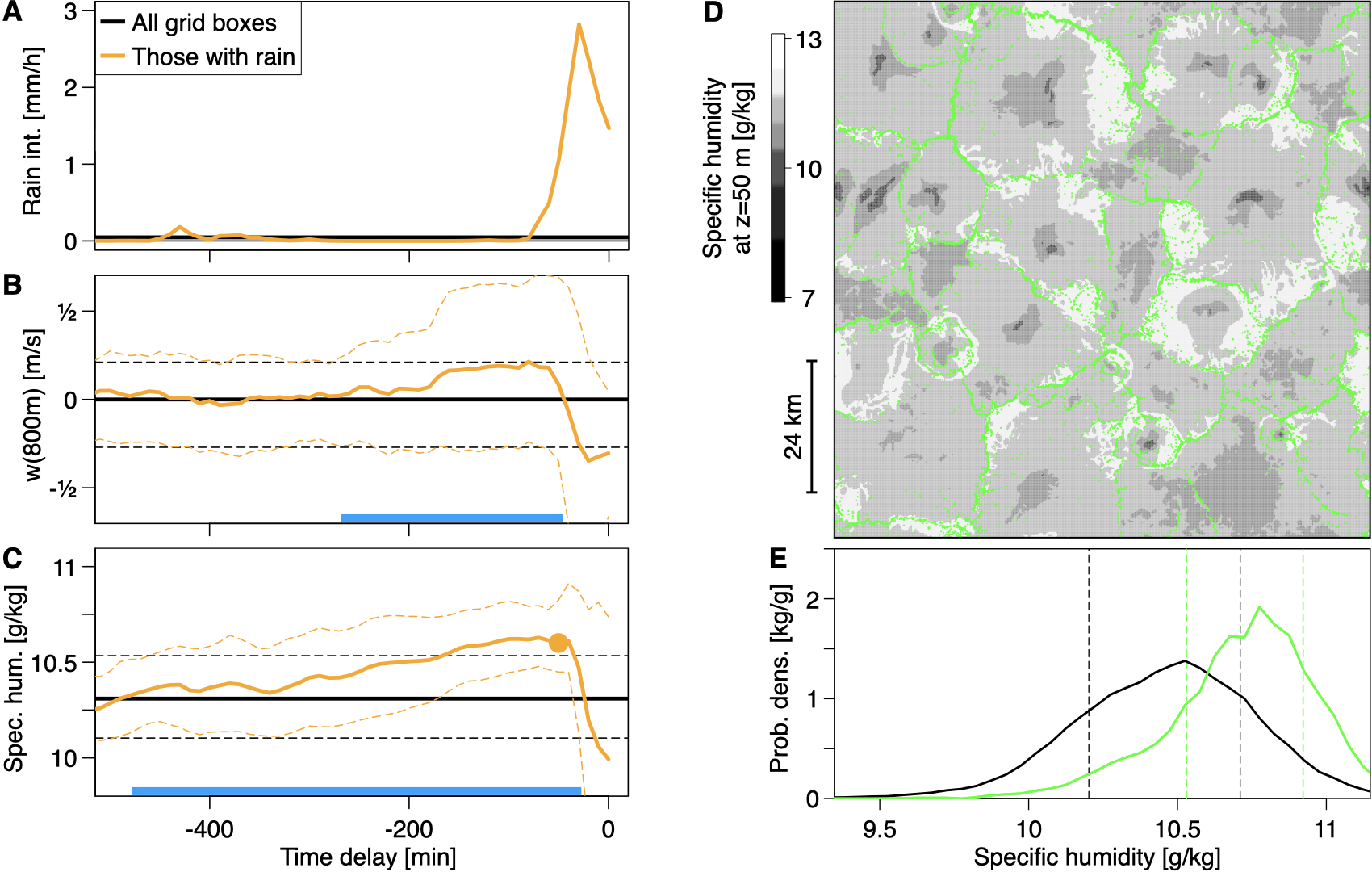}
\caption{{\bf Identifying cold pool collisions}. 
{\bf (A)} Rain intensity conditionally averaged over all grid boxes with rainfall at $t$ = 0 (orange) and domain mean rainfall (black) for the simulation Evap=1. 
{\bf (B)} Analog, but for vertical velocity near the cloud base ($w$(800 m)). 
The domain average is zero throughout.
Thin lines mark corresponding 20th and 80th percentiles.
Note the pronounced peak, corresponding to convective updrafts, as is expected before rain onset, as well as the dip near $t$ = 0, corresponding to CP associated downdrafts.
{\bf (C)} Analogous, but for near-surface specific humidity, $q_v$($z$ = 50 m). 
Note the relatively long buildup of humidity before rainfall onset. 
The blue bar highlights the time during which updrafts exceed the domain average.
{\bf (D)} Near-surface specific humidity field, $q_v$($z$ = 50 m).
Tracer particles (green) visually lie at locations of CP gust fronts. 
The blue bar highlights the time during which specific humidity exceeds the domain average.
{\bf (E)} Histograms of $q_v$(50 m) for all data shown in panel D (black curve), and only the gust front positions (the green points in panel D).
}
\label{fig:tracers}
\end{figure*}

\noindent
{\bf New convective events are initiated in the vicinity of cold pool collisions.}
What is then the specific role of CPs in maintaining domain-wide convection?
To explore this, first consider locations of rainfall at a particular time step of Evap=1 (Fig.~\ref{fig:tracers}A), the associated cloud-base vertical velocity (Fig.~\ref{fig:tracers}B), and specific humidity (Fig.~\ref{fig:tracers}C). 
Updrafts form shortly before the onset of rainfall, whereas specific humidity becomes elevated earlier --- in line with the analysis of RCE simulations, where a considerable buoyancy build-up before any subsequent convective event was reported.\cite{fuglestvedt2020cold}
Second, we determine gust front loci using CP tracer particles \cite{haerter2019circling,henneberg2020particle} (Fig.~\ref{fig:tracers}D).
Many of these tracers are located at the intersection between two CPs.
It is visually apparent that such loci coincide with enhanced humidity (Fig.~\ref{fig:tracers}D).
To quantify that tracers lie in regions of pronounced updrafts, we verified that the $q_v$(50 m)-histogram for all loci with tracers is shifted to markedly positive values (Fig.~\ref{fig:tracers}E). 
The histogram of peak humidity during rain event buildup (peak highlighted in Fig.~\ref{fig:tracers}C) shows a shift towards elevated values.
Collecting, as a comparison, the specific humidity at CP gust fronts (Fig.~\ref{fig:tracers}E), it is found that this histogram is similarly shifted to moister values.
In summary, loci of CP collisions do provide the positive humidity anomalies typical of subsequent convective events.

\begin{figure}[htb]
\centering
\includegraphics[width=0.85\linewidth]{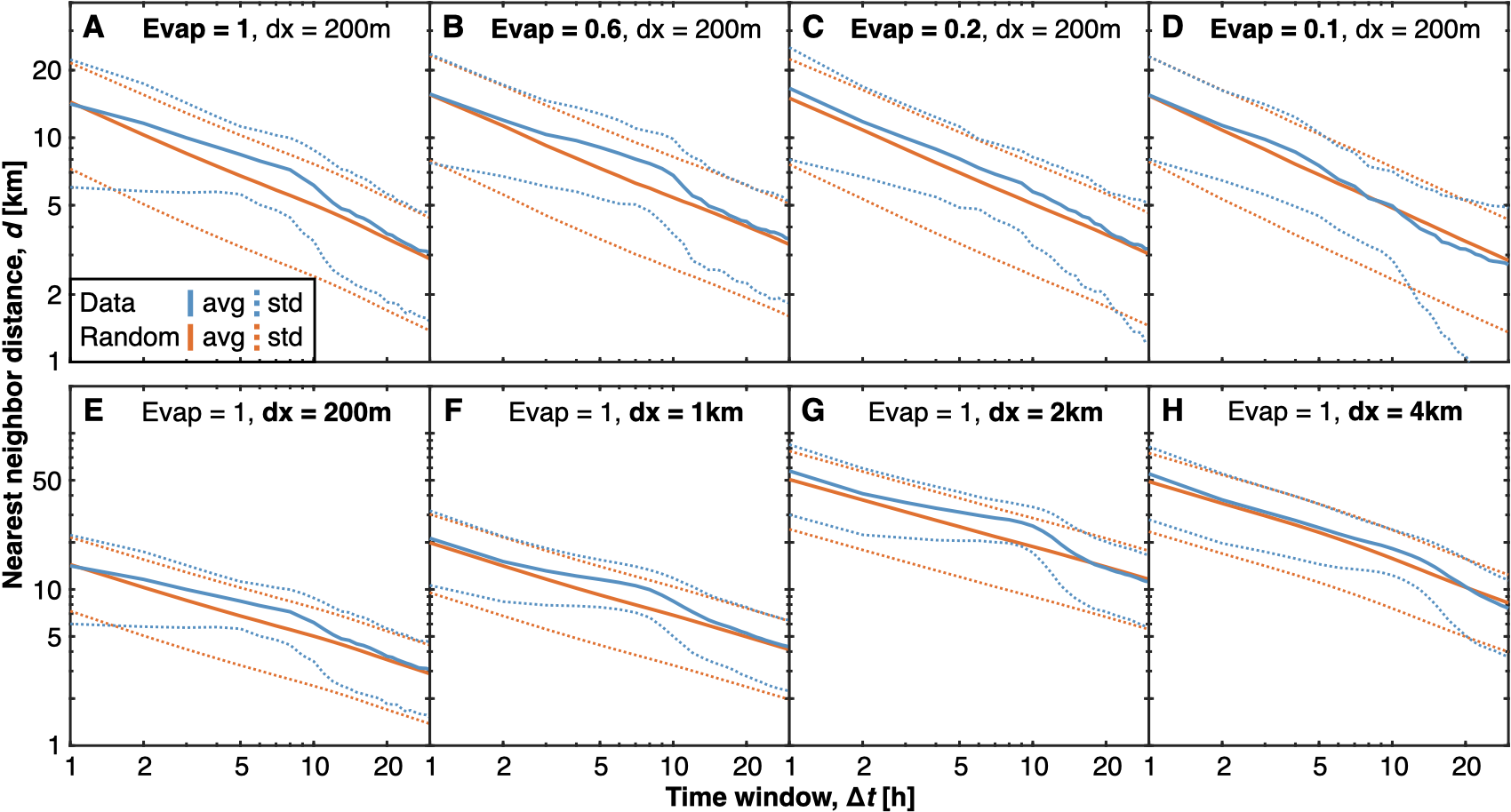}
\caption{{\bf Generation time and an effective minimal radius, $\mathbf{R_{min}}$}. 
{\bf (A--D)} The average distance ($d$) between rain events occurring within the first 12h past precipitation onset and their nearest neighbor rain event (blue) occurring within a time window ($\Delta t$) for varying evaporation rates (Evap). 
We contrast that to a control where the same number of rain events are located randomly (red), given as the average distance $d$ of the probability density function $f(d)=2n \pi d (1-\pi d^2/L^2)^{-1+n}/L^2$ where $L$ is the domain length and $n$ is the mean number of rain events during $\Delta t$. 
The dotted lines mark the standard deviations. 
%The vertical lines indicate average numbers.
%The dashed line marks an accumulated frequency of 0.1.
More rain events are included for larger time windows, causing the distances to be smaller. 
In (A), note the lack of events within 5--6 km for up to 8h. %$\Delta t$ = 6h. 
{\bf (E--H)} Analogous, but for varying horizontal resolutions ($dx$). 
Note the inhibitory distance increases for coarser resolutions without changing the time scale at which it occurs.
}
\label{fig:Rmin}
\end{figure}

\noindent
{\bf New deep convective events are initiated at a specific distance away from earlier events.}
To quantify a possible suppression effect caused by a present rain cell's CP on subsequent cells forming within the surroundings, we examine whether rain events after the initial rain onset are %spatially located in a way that favors self-aggregation. 
%Specifically, we ask whether the events are 
spaced randomly or not. 
A non-random spacing would imply either suppression (larger distance) or activation (smaller distance), whereas a random spacing would speak against a direct spatial influence on subsequent rain cell formation.
We thus identify all rain events within the first 12h after rain onset \cite{moseley2013probing} allowing us to compare non-aggregating simulations with aggregating simulations (day 1 in Fig.~\ref{fig:specific_humidity}). 
We measure each rain cell's distance to it's nearest rain event occurring within a time window $\Delta t$. 

For the control simulation (Fig.~\ref{fig:Rmin}A), we find an inhibitory effect causing the nearest neighbor distance to be $>5$ km for up to 8h. In contrast, it would linear decay on a log-log plot if events were distributed randomly in space ({\it compare} orange and blue curves in Fig.~\ref{fig:Rmin}). 
We refer to this distance as $R_{\text{min}}\approx$ 6 km and explain it by CPs being too negatively buoyant to initialize new convective cells within this distance \cite{drager2017characterizing,fournier2019tracking}.
When increasing the time window of included events from 8--12 hours, we find that this suppression effect disappears; that is, the distribution function matches the random one. 
On this time scale, the CPs associated with two rain events have time to grow larger than $R_{\text{min}}$, collide, and trigger the formation of a new, closer, rain event belonging to the subsequent generation. 
We, therefore, interpret this time scale as the generation time of one CP. 
We find this time scale unchanged across simulations, while the spatial scale increases for coarser horizontal resolutions (Fig.~\ref{fig:Rmin}). 

\begin{figure*}[htb]
\centering
\includegraphics[width=0.85\linewidth]{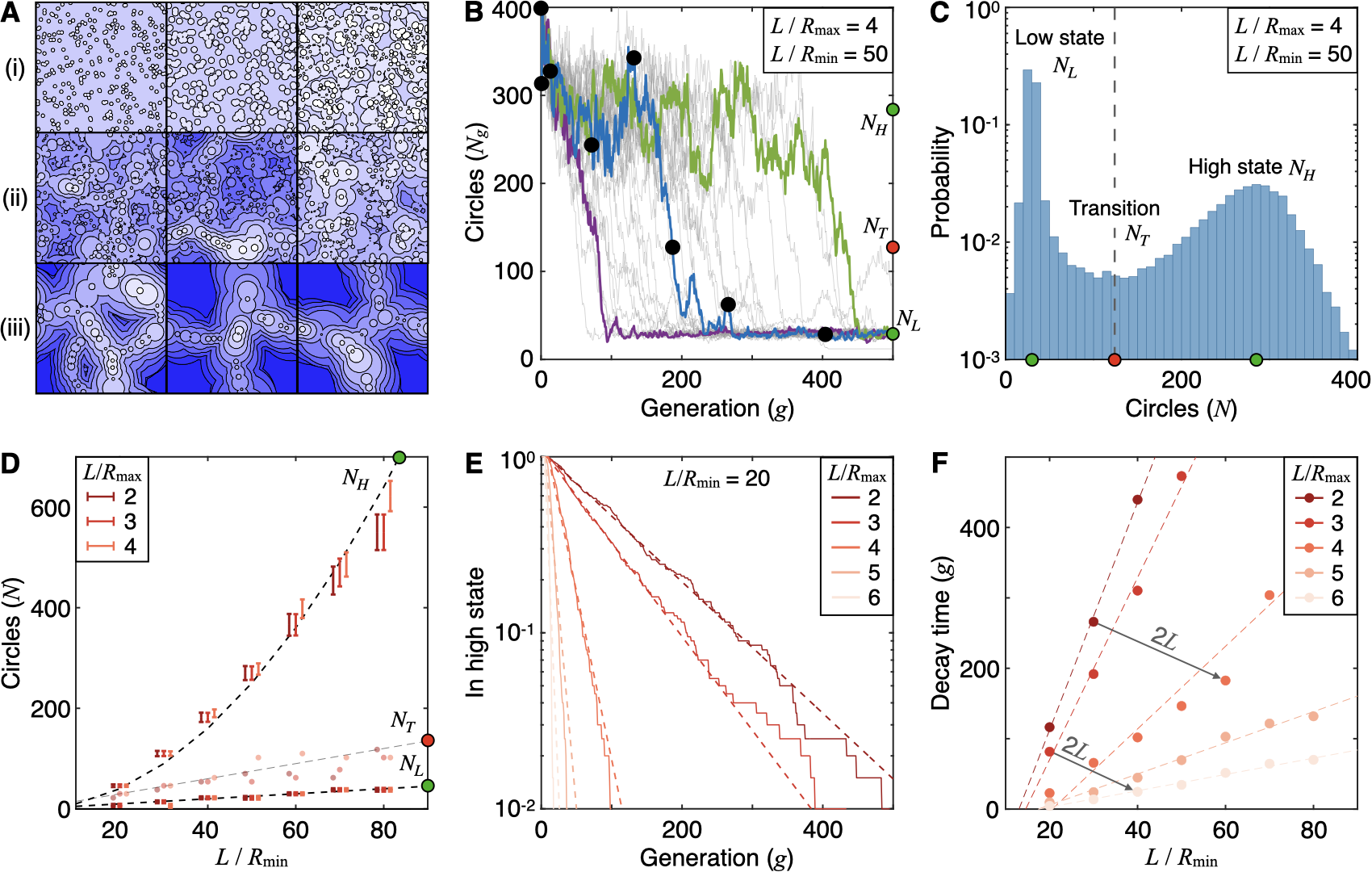}
\caption{{\bf Circle model.} {\bf (A)} $N_1$ randomly located circles representing cold pools expand radially with equal and constant speed on an $L \times L$ domain. 
Any two colliding circles of radii $R_{\text{min}}<r_i, r_j<R_{\text{max}}$ initiate a new expanding circle at their collision point ({\it Details:} Methods). 
Within each sub-panel, circles of the same color belong to the same generation and white (dark blue) areas correspond to the most recent (historic) generations in the respective sub-panel. 
{\bf (B)} The number of circles ($N_g$) per generation ($g$) as 30 simulations evolve. Three runs are highlighted. The blue curve represents the simulation in (A), and the black dots indicate the snapshots' time. 
Note the existence of two qualitatively distinct states of high and low $N_g$.
{\bf (C)} The distribution of circles in all generations pooled together for 200 runs. 
In (A--C), $L/R_{\text{min}}$ = 50 and $L / R_{\text{max}}$ = 4. 
In (B--D), the two green circles mark the low and the high states while the red circle marks the transition point.
{\bf (D)} The number of circles in the low $N_L$ and high $N_H$ states, and at the transition point $N_T$ for varying $L/R_{\text{max}}$ and $L/R_{\text{min}}$. 
{\bf (E)} The fraction of runs in the high state as generation number for $L/R_{\text{min}}$ = 20. 
Note the logarithmic vertical axes.
{\bf (F)} The characteristic decay time as a function of $L / R_{\text{max}}$ and $L/R_{\text{min}}$. 
Note that higher $L$, higher $R_{\text{min}}$, or lower $R_{\text{max}}$ result in faster decay.}
%{\bf (A)} (i): $N$ = 60 randomly located circles representing cold pools expand with equal and constant speed on a 2D domain of $L$ = 100 $\times$ 100 km. Two colliding circles of radii $R_{\text{min}}$ = 5 km $<r_i, r_j<R_{\text{max}}$ = 30 km initiate new expanding circles at their collision point.
%(ii): The model continues and reaches a meta-stable state with homogeneous activity.
%(iii): Impurities in the circles' location cause a non-active region to form.
%The colors indicate the circles' generation.
%{\bf (B)} The number of circles as simulations with various domain sizes $L$ and maximum radii $R_{\text{max}}$ evolve.
%One generation = 8 hours and $R_{\text{min}}$ = 4 km. 
%{\bf (C)} The same with fixed $R_{\text{max}}$ = 50 km. 
%Note the existence of two states: a meta-stable state with saturating activity, $N\propto L^2/R_{\text{min}}^2$ and an aggregating state with low activity, $N\propto L/R_{\text{min}}$.
%Note that simulations with lower $R_{\text{max}}$, higher $L$, or higher $R_{\text{min}}$ fall to the aggregating state faster.
\label{fig:circle_model}
\end{figure*}

\noindent
{\bf A simple mathematical model captures the onset of self-aggregation.}
To understand the role of CP collisions, we introduce a model consisting of growing and colliding circles. 
The circles represent the gust fronts of CPs and they are initialized uniformly at random inside the model domain. 
The circle radius, $r$, initially set to zero, increases linearly until $r=R_{\text{max}}$, hence the expansion velocity, $v(r)$, is a step function $v(r) = v_0\theta(R_{\text{max}}-r)$ where $v_0$ is a constant and $\theta(R_{\text{max}}-r)=1$ for $r<R_{\text{max}}$, otherwise $\theta=0$. 
After a given circle reaches $R_{\text{max}}$, it has no further effect and is removed.
When two circles meet, both having their radii lie between $R_{\text{min}}$ and $R_{\text{max}}$, they instantly produce a new circle at the first point of intersection.
The reasoning is that in RCE, most new rain cells result from thermodynamic pre-conditioning near the gust front collision lines (Fig.~\ref{fig:tracers} and \citeauthor{fuglestvedt2020cold}, \citeyear{fuglestvedt2020cold}), and the delay between the collision time and the initiation of the resultant rain cell is so large (typically several hours) that direct forced lifting can be ruled out. 
%The simplest 2-CP model includes a minimum radius, $R_{\text{min}}$, that both circles need to trigger new events. 
In line with the findings in Fig.~\ref{fig:Rmin}, CPs with $r<R_{\text{min}}$ are considered too negatively buoyant to initialize new CPs \cite{drager2017characterizing,fournier2019tracking}. 
%We further incorporate the maximum radius, $R_{\text{max}}$, reached by the CPs. 
However, circles may collide with multiple other circles until they reach $R_{\text{max}}$ beyond which they have no role.

Mathematically, the collision dynamics allow us to categorize circles into {\it generations}: 
the initially seeded circles constitute generation one (panels 1--2 in Fig.~\ref{fig:circle_model}A,i where all circles are smaller than $R_{\text{min}}$). 
Each collision is between circles belonging to the same generation $g$, and the resultant new circles belong to the subsequent generation $g+1$.
$R_{\text{min}}>0$ prevents singularities, that is, infinitely rapid replication.
Given $R_{\text{min}}=0$, an initial random generation-one population $N_1$ of circles would yield $N_2=2\,N_1$ ({\it Details:} Supplement), and subsequent growth of $N_g$ vs.~$g$ would be nearly exponential.
For $R_{\text{min}}>0$ circles must grow beyond $R_{\text{min}}$ and only then replicate to yield generation-two events (note the different color shadings in Fig.~\ref{fig:circle_model}A,i, third panel). 
%In Fig.~\ref{fig:circle_model}A, we present the spatial dynamics of a simulation. %, where self-aggregation occurred, characterized by the presence of circle-free sub-regions (Fig.~\ref{fig:circle_model}A,iii) and the visual similarity with the numerical experiment in Fig.~\ref{fig:specific_humidity}C--D. 
%Stage (i) shows the random initialization of rain cells ({\it termed} generation 1) and the first replication, leading to generation 2 events. 
In Fig.~\ref{fig:circle_model}A,ii new circles of various generations are initiated throughout the domain with no obvious patterning. 
However, during Fig.~\ref{fig:circle_model}A,iii a separation into a circle-filled (convecting) and a circle-free (non-convecting) sub-region occurs. 
Note the visual similarity with the numerical experiment in Fig.~\ref{fig:specific_humidity}C--D. 

The number of circles $N_g$ in all simulations eventually drops from an initial high-$N$ state to one with low $N$ (Fig.~\ref{fig:circle_model}B).
The histogram of $N$, which is bimodal, confirms the notion of two distinct meta-stable states (Fig.~\ref{fig:circle_model}C). 
Given this finding, we explore how the two states depend on the independent model parameters $L/R_{\text{min}}$ and $L/R_{\text{max}}$. 
We find that the low-$N$ state scales as $N_L = L/(2R_{\text{min}})$, whereas the high-$N$ state scales as $N_H = L^2/(10R_{\text{min}}^2)$, both independent of $R_{\text{max}}$ (Fig.~\ref{fig:circle_model}D). 
The transition point occurs at $N_T \approx 1.5L/R_{\text{min}}$
The linear scaling $N_L\sim L$ is commensurate with band-like, one-dimensional, structures (compare Fig.~\ref{fig:circle_model}A,iii and Fig.~\ref{fig:specific_humidity}C--D on day 2--4), whereas $N_H\sim L^2$ is in line with two-dimensional organization. 
By fitting the fraction of simulations in the high-$N$ state to an exponential function (Fig.~\ref{fig:circle_model}E), we show that a characteristic time exists when the simulations decay to the low state. 
Thereby, we find that the circle model predicts decreasing $R_{\text{max}}$, increasing $L$, or increasing $R_{\text{min}}$ speed up the characteristic time when the transition occurs (Fig.~\ref{fig:circle_model}F). 
Decreasing $R_{\text{max}}$ is in correspondence with the results presented in Figs.~\ref{fig:specific_humidity}--\ref{fig:Rmax}. 
Increasing $L$ has previously been reported to facilitate self-aggregation \cite{bretherton2005energy,muller2015favors}. The literature also matches that coarser horizontal resolution favors self-aggregation given our result in Fig.~\ref{fig:Rmin}E--H where we show that coarser horizontal resolution results in increased $R_{\text{min}}$ \cite{hirt2020cold,yanase2020new}.

\section{Discussion and Conclusion}
Convincing theories have been proposed for Turing-like coarsening of the RCE atmosphere into moist and dry sub-regions \cite{bretherton2005energy,craig2013coarsening,emanuel2014radiative}.
Such reaction-diffusion systems describe local positive feedback, e.g., of moisture on itself. 
In contrast, we here explicitly model the non-local, two-particle interaction resulting from interacting cold pool (CP) gust fronts --- competing with the well-studied local moisture feedback.
The notion that CP collisions form new convective events is well documented \cite{purdom1976some,weaver1982multiscale,torri2019cold} and addressed in toy models \cite{boing2016object,haerter2019convective}. 
Our study investigates how the CP radius influences CSA and builds on the finding that the total rain cell number is approximately constant across simulations (Fig.~S3).
Indeed, relatively constant rain cell numbers (Fig.~S3) and rain intensities (Fig.~S2B) are supported by radiation constraints on precipitation in RCE \cite{held2006robust}.
Our circle model implicates that large CPs, as formed by pronounced rain evaporation, become space-filling where the whole domain is filled by CPs and there is a connected patch through the domain among CPs from the same generation. 
From hexagonal close-packing, a lower radius bound for space-filling would be $R_{\text{max}}>L(3\sqrt{3}N)^{-1/2}\approx 5.7\,km$, where $L=92\,km$ is the side length of the simulation domain and $N\approx 50$ is the number of CPs per generation (Fig.~S3).
For radii smaller than this, areas emerge that cannot be reached by newly initialized circles --- a gap results, and the transition to CSA sets in.
When (realistically) departing from perfect close-packing, the required value of $R_{\text{max}}$ may lie somewhat higher --- commensurate with our findings (Fig.~\ref{fig:Rmax}) and the transition to CSA between Evap=.6, where $R_{\text{max}}\approx$ 12 km, and Evap=.2, where $R_{\text{max}}\approx$ 8 km. 

%Each newly formed CP will inevitably collide with pre-existing gust fronts from other CPs --- thus allowing for convective triggering at the corresponding interfaces and maintaining rain events being distributed homogeneously through the domain.
%Smaller CPs, unable to fill up space, will only occasionally collide with other CPs.
%CP replication, therefore, remains low, and self-aggregation will ensue.

%On a unit domain, our model reduces the impact of CP interaction onto convective organization to two parameters: two parameters: a minimal ($R_{\text{min}}$) and a maximal ($R_{\text{max}}$) circle radius, which are important to determine the organization state. 
%We extract these from a suite of simulations and justify them from the physics of CPs: 
%$R_{\text{min}}$ results from CPs showing strongly negative buoyancy at initial stage caused by dry downdrafts and $R_{\text{max}}$ results from CP dissipation. 
Our model simplifies CP expansion by assuming a step function $v(r) = v_0\theta(R_{\text{max}}-r)$. 
In reality, CPs initially grow quickly and their speed of spreading decreases gradually over the course of a few hours (Fig.~\ref{fig:Rmax}B) \cite{grant2016cold,grant2018cold}. 
Introducing a smoothly varying gust front speed into our model would require a time-dependent expansion speed factor and
%in Equations \ref{eq:circ1}--\ref{eq:circ2}. 
a numerically-approximate approach might then be more practicable \cite{haerter2019circling}. 
The presented model further does not reach a final, fully-aggregated state, where a small fraction of the domain intensely convects indefinitely. 
This sustained activity might be obtained by adding spatial noise (displacing new circles slightly away from the exact geometric collision point) and systematically increased triggering probabilities for decreased overall rain area \cite{haerter2019convective}. 
Extensions could include explicit incorporation of the "super-CP" \cite{windmiller2019convection} and radiatively driven CP \cite{coppin2015physical,yanase2020new}, constituting the two components of the final large-scale circulation. 
%which forms under the final large scale circulation and transports moisture from the dry to the moist sub-region. 
This model extension would allow triggering events at the edges of the intensely-convecting supercell due to convective CPs colliding with the opposing radiatively driven CP. 
Such circulation feedbacks may well be essential in stabilizing the final steady state but may not be required to develop the first dry patches and their initial growth, which we have focused on in this work.

\section*{Acknowledgments}
\noindent
We thank Steven J.~B{\"o}ing, Cathy Hohenegger, and the Atmospheric Complexity Group at the Niels Bohr Institute for discussions.
The source code for the mathematical model is available at \url{https://github.com/SilasBoyeNissen/How-weakened-cold-pools-open-for-convective-self-aggregation}.
SBN acknowledges funding through the Danish National Research Foundation (grant number: DNRF116). 
JOH gratefully acknowledges funding by a grant from the VILLUM Foundation (grant number: 13168) and the European Research Council (ERC) under the European Union's Horizon 2020 research and innovation program (grant number: 771859). 
We acknowledge the Danish Climate Computing Center (DC3) and thank Roman Nuterman for technical support.
The authors gratefully acknowledge the German Climate Computing Centre (Deutsches Klimarechenzentrum, DKRZ).

\section*{Author Contributions}
\noindent
JOH ran and processed the large-eddy simulations, tracked the rain cells and cold pools, and contributed to the manuscript. 
SBN analyzed the simulation data, developed, implemented, analyzed the circle model, and drafted and revised the manuscript.

\section*{Competing Interests}
\noindent
The authors declare no competing interests.

\bibliography{main.bib}

\newpage

\renewcommand{\theequation}{S\arabic{equation}}
\renewcommand{\thesection}{S\arabic{section}}
\renewcommand{\thefigure}{S\arabic{figure}}
\renewcommand{\thetable}{S\arabic{table}}

\setcounter{equation}{0}
\setcounter{figure}{0}
\setcounter{page}{1}
%\resetlinenumber

\section*{How weakened cold pools open for convective self-aggregation}
\begin{center}
Silas Boye Nissen and Jan O.~Haerter
\end{center}

\subsection*{Supplementary Information}\label{sec:supp}
\begin{figure*}[htb]
\centering
\includegraphics[width=0.35\linewidth]{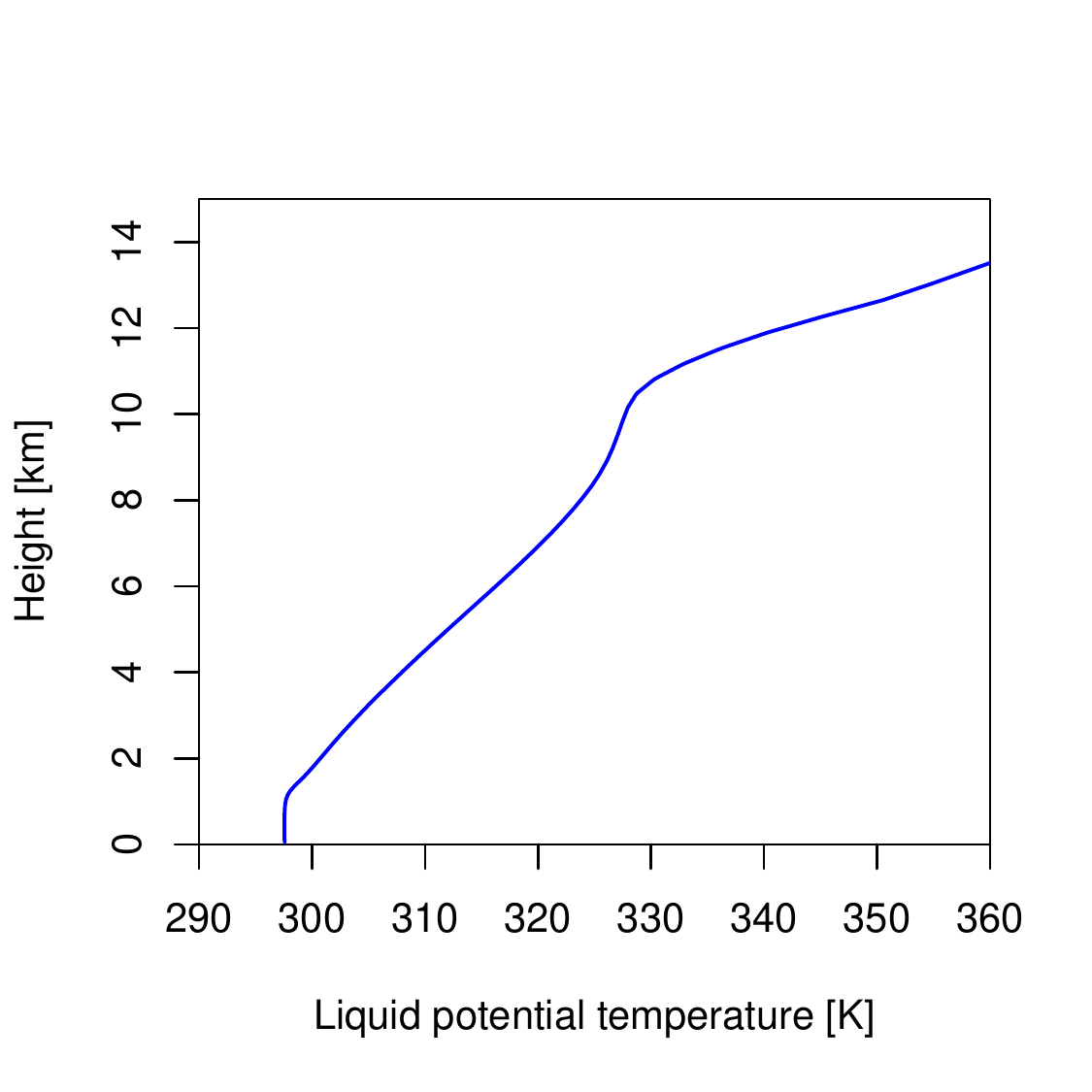}
\includegraphics[width=0.35\linewidth]{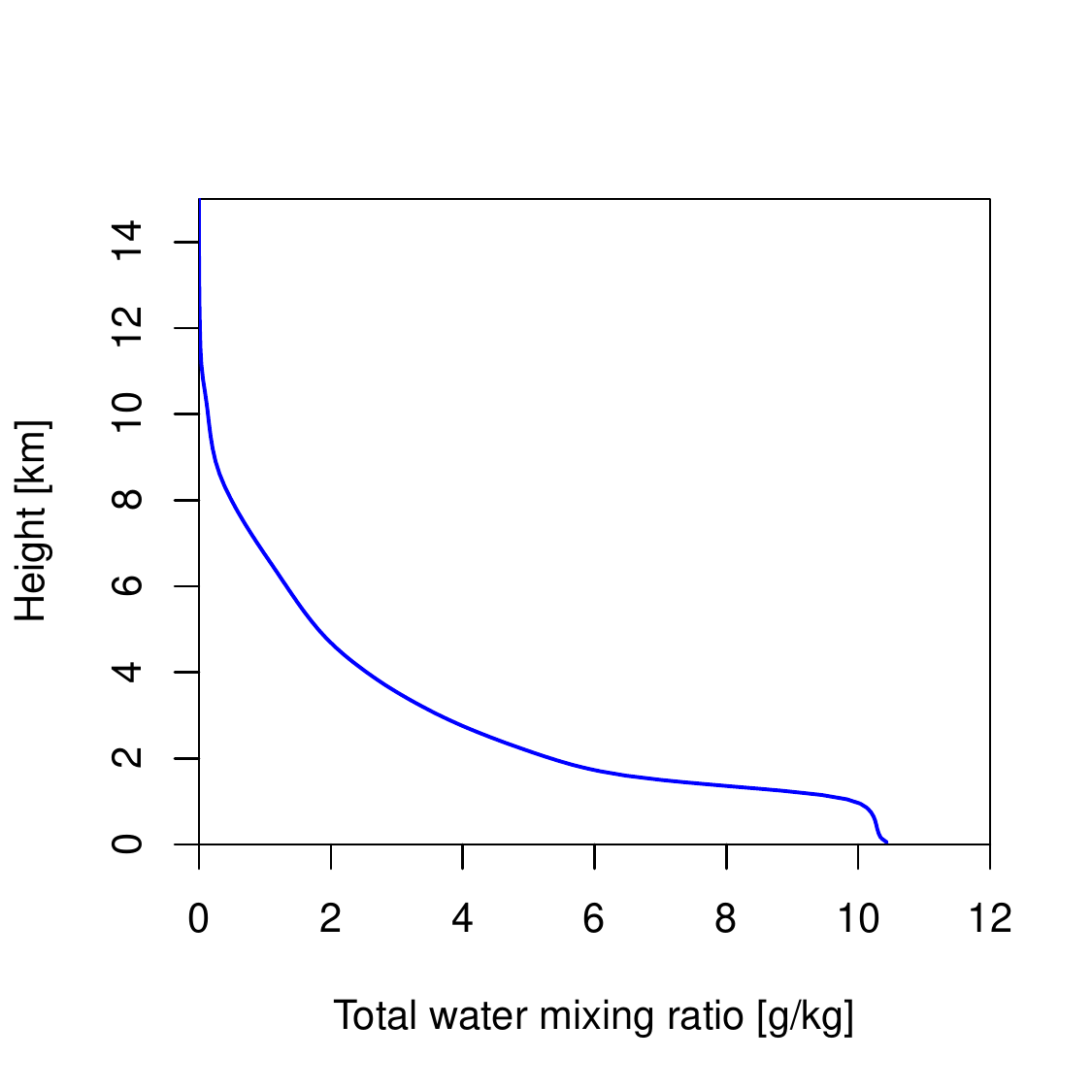}
\vspace{-.5cm}
\caption{{\bf Initial condition.} 
Vertical profiles for temperature (left) and moisture (right) used as the initial condition for all numerical experiments. 
}
\label{fig:InitialConditions}
\end{figure*}

\begin{figure*}[htb]
\centering
\begin{overpic}[trim={0 0cm 0 0},clip,width=0.3\textwidth]{dummy.pdf}
\put(-83,0){\includegraphics[trim={0cm 0cm 0cm 0cm}, clip, width=0.35\linewidth]{
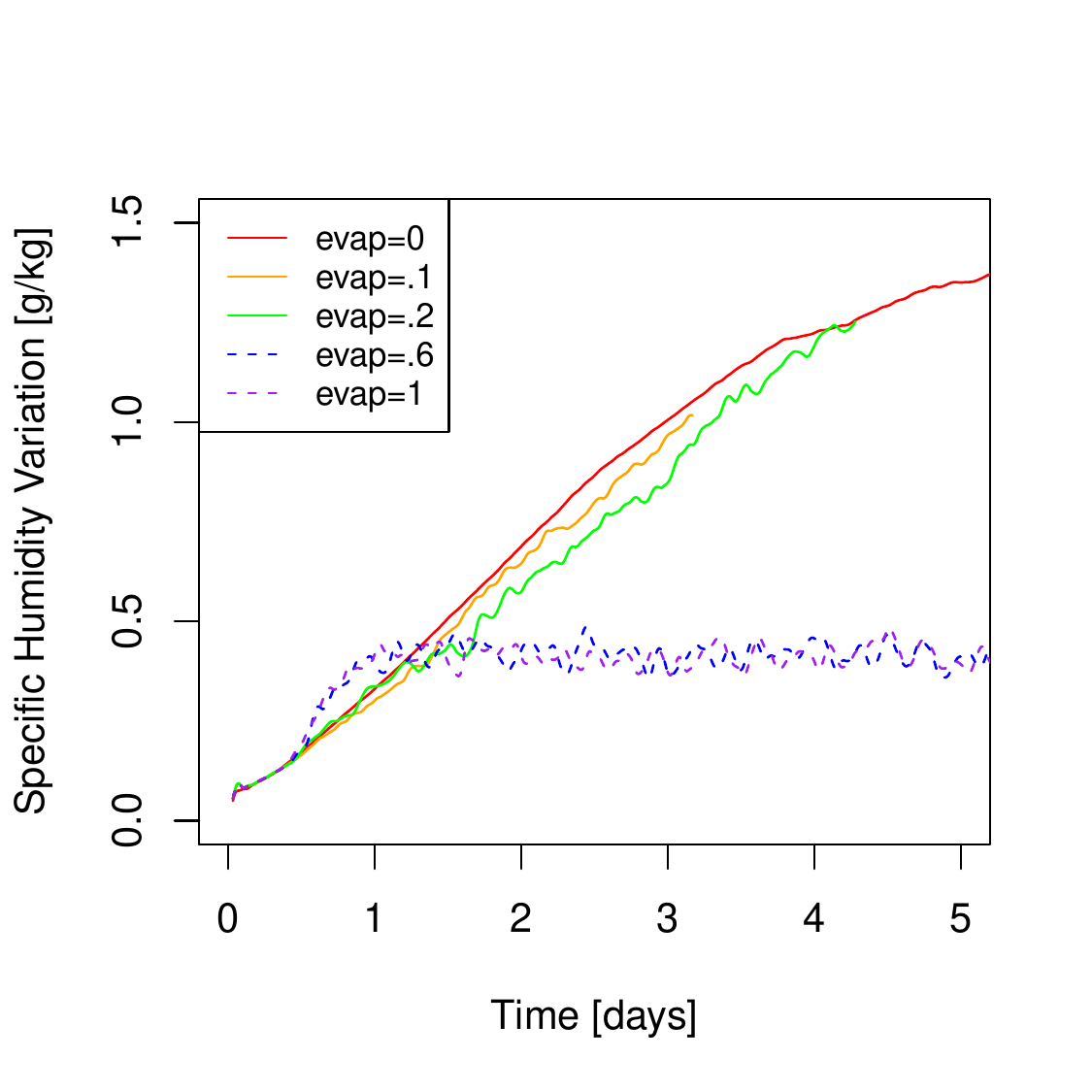}}
\put(-4,0){\includegraphics[width=0.35\linewidth]{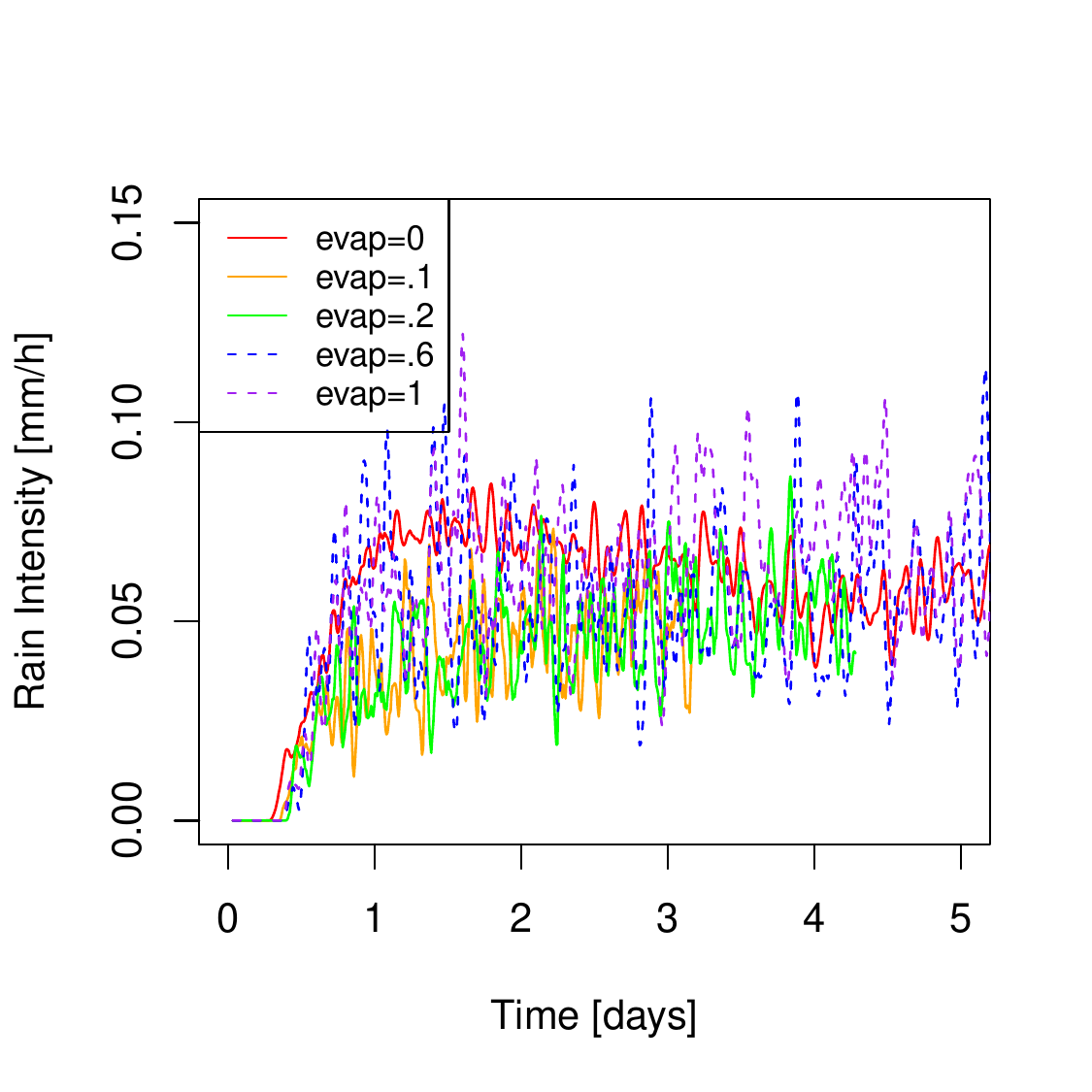}}
\put(75,0){\includegraphics[width=0.35\linewidth]{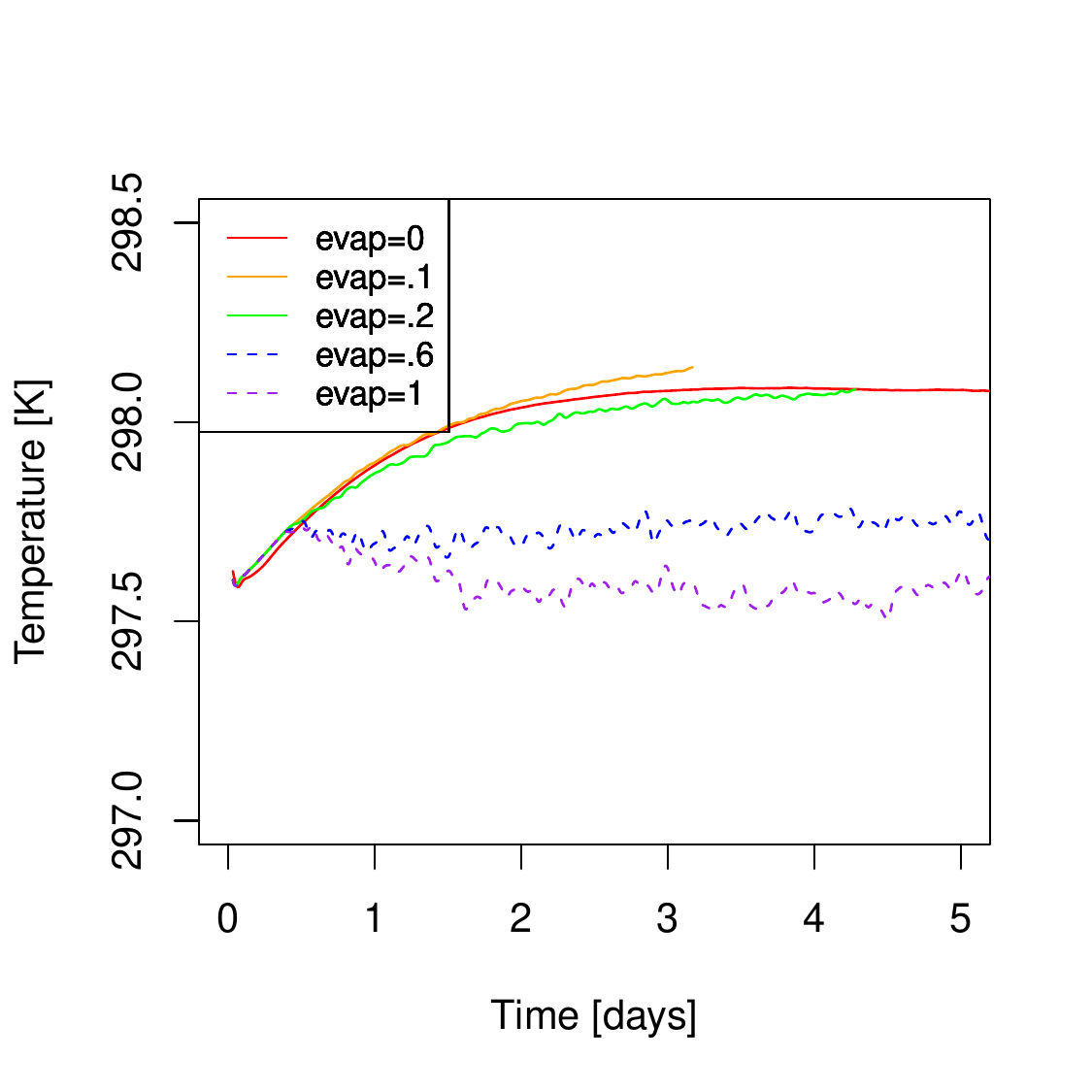}}
\put(-83,70){\large \bf A}
\put(-4,70){\large \bf B}
\put(75,70){\large \bf C}
\end{overpic}
\caption{{\bf Humidity variation, domain mean rainfall, and temperature}. 
{\bf (A)} Each curve ({\it see} legend for line-style) was computed from the difference $\Delta q(t)\equiv q_{v,75}(t)-q_{v,25}(t)$, where the subscript numbers denote the respective percentiles of the specific humidity for each simulation and the argument $t$ denotes the specific simulation output time step.
The continued increase for Evap $\in\{0,.1,.2\}$ signals the onset of self-aggregation for those cases.
{\bf (B)} Domain-mean rainfall for each simulation over time. 
{\bf (C)} Domain-mean temperature for each simulation over time.
We attribute the kink for the dis-aggregated simulations (Evap=.6 and Evap=1) near $t$ = .5 day to the onset of rainfall.
}
\label{fig:q_timeseries_max-min}
\end{figure*}

\begin{figure*}[htb]
\centering
\includegraphics[width=0.4\linewidth]{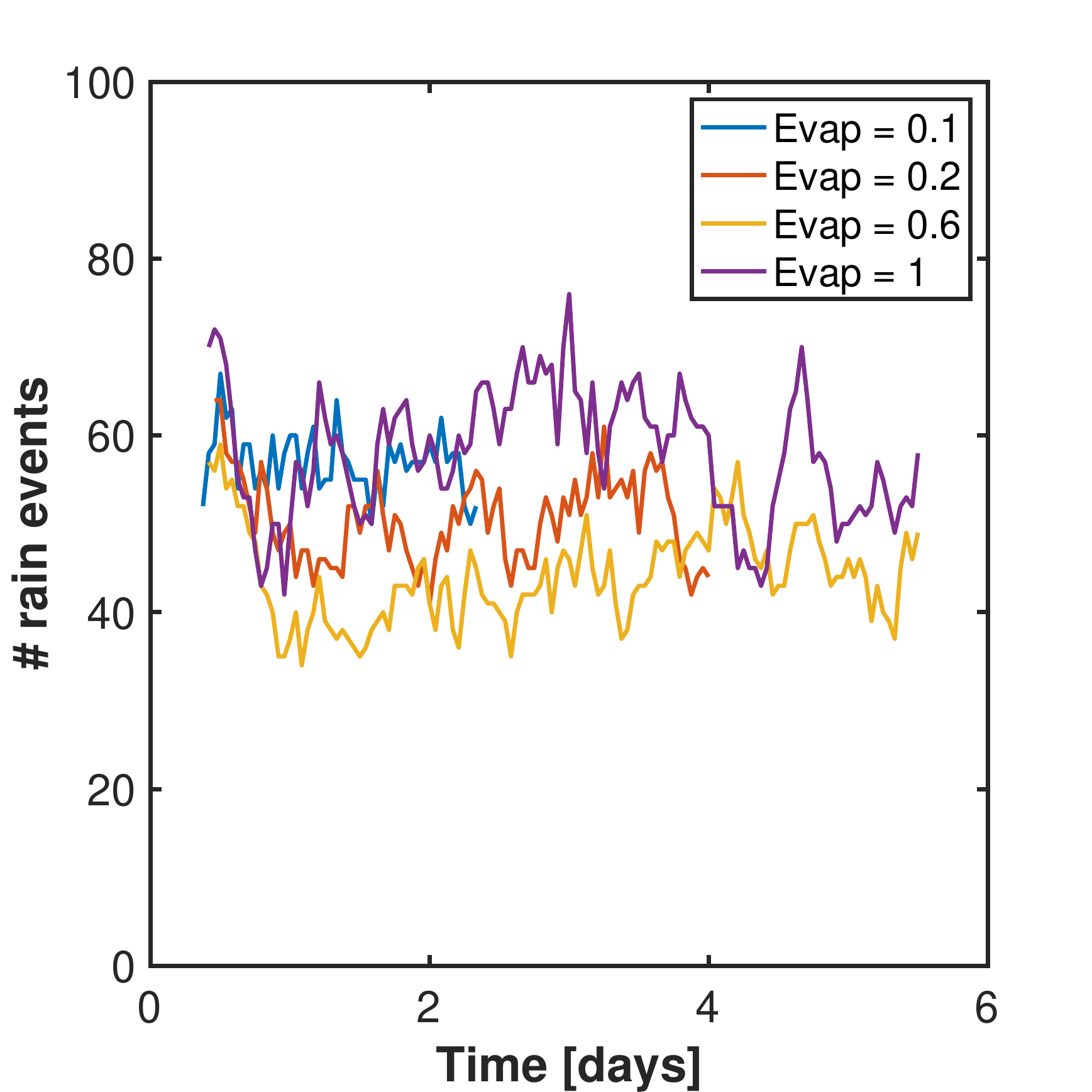}
\caption{{\bf Timeseries of the number of rainfall tracks.} 
Track counts were computed within a 6-hour running time window.}
\label{fig:NoEvents}
\end{figure*}

\noindent
Here, we analytically find the average number of circle collisions for neighbors in a system with randomly positioned cells, and we provide supplementary figures, including the initial conditions for the large-eddy simulations (Fig.~\ref{fig:InitialConditions}), the time development of the domain-mean specific humidity variation, rain intensity, and temperature (Fig.~\ref{fig:q_timeseries_max-min}), as well as the number of rainfall tracks (Fig.~\ref{fig:NoEvents}).

\begin{figure*}
\centering
\includegraphics[height=5cm, trim=0cm 0cm 0cm 0cm, clip]{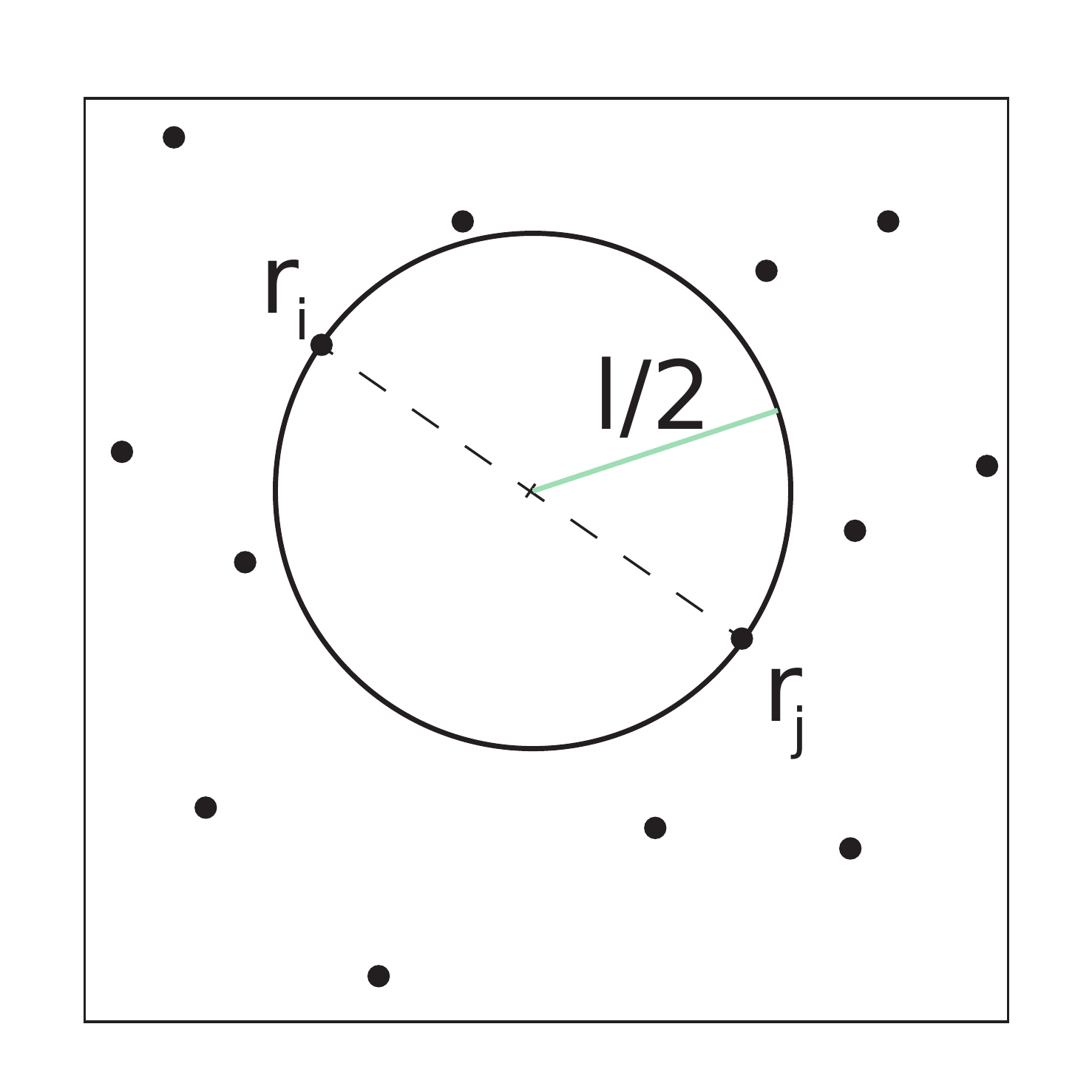}
\caption{{\bf A line-of-sight connection.} Schematic illustrating points in a 2D domain. Two points, ${\mathbf r_i}$ and ${\mathbf r_j}$, separated by a distance $l$ have a line-of-sight connection given that no points are located inside the circle of radius $l/2$ with the two points on its rim. 
In mathematics, this concept is also known as Gabriel neighbors.}
\label{fig:los}
\end{figure*}

\noindent
{\bf Direct circle collisions for random seeding.}\label{sec:los}
We seek to compute the number of {\it direct} circle collisions for an initial random population of circle centers, with circles emerging synchronously and at equal speed from all circle centers.
By {\it direct} we mean, that the collision takes place along the line connecting the two circle centers involved.
Finding the number of such collisions is equivalent to finding the number of {\it line-of-sight} connections (also known as a Gabriel connection) among these points (Fig.~\ref{fig:los}).

Consider $N$ points randomly seeded in a total area $A=\pi R^2$. 
A line-of-sight connection between any two points $\mathbf{r_i}$ and $\mathbf{r_j}$ at distance $l\equiv |\mathbf{r_i}-\mathbf{r_j}|$ exists if there are no points located inside the circle of radius $l/2$ centered at $(\mathbf{r_i}+\mathbf{r_j})/2$ (Fig.~\ref{fig:los}). This is the type of connection that we require for any two colliding circles to initialize the growth of a new circle in Fig.~5.
One must further consider the probability of finding two points at distance $l$. 
For this purpose, define the density of points as $\rho\equiv N/A=N/\pi R^2$. 
Now consider the infinitesimal area $a(l)\equiv 2\pi\,l\,dl$ between two circles of radii $l$ and $l+dl$. 
The number of points contained in this area is
\begin{equation}
    n(l)dl = \rho\,a(l)= \frac{2Nl}{R^2}dl.
\end{equation}

For any two points at a given distance $l$, we now consider the probability $p(l)$, that none of the remaining $N-2$ points lie within the circle of radius $l/2$:
\begin{equation}
    P(l) = P_0^{N-2}=\left( 1-\frac{\pi(l/2)^2}{\pi R^2}\right)^{N-2},
\end{equation}

where
\begin{equation}
    P_0 = 1-\frac{\pi(l/2)^2}{\pi R^2}
\end{equation}

is the probability that any single point is not inside the area enclosed by a circle of radius $l/2$. 
Now the total number of expected line-of-sight (LOS) connections for a fixed given point to any of the other points can be computed:
\begin{eqnarray}
    N_{\text{LOS}} &=& \int_0^R dl\;n(l) P(l)\\
    &=& \int_0^R dl\frac{2Nl}{R^2}\left( 1-\frac{l^2}{4R^2}\right)^{N-2}\\
    &=& \frac{4(1-\left( \frac{3}{4}\right)^{N-1})}{1-N^{-1}},
\end{eqnarray}
\noindent
which gives $\lim_{N\rightarrow\infty}N_{\text{LOS}}$ = 4. Hence, when repeating for all $N$ and avoiding double-counting of connections, one obtains that there are $2N$ line-of-sight connections when going from generation 1 to 2 for $R_{\text{min}}=0$ and $R_{\text{max}}=\infty$.

\end{document}